\documentclass[12pt]{article}

\usepackage{graphics}
\usepackage[dvips]{graphicx}

\begin{document}

{\footnotesize jcis@epacis.net}

\begin{center}

{\bf Two semi-automated computational approaches \\
for spectroscopic analysis of stellar photospheres:\\
detailed characterization at high resolution\\
and abundance determination at medium resolution\\}
\bigskip


{\small Andr\'e Milone$^a$\footnote{E-mail Corresponding Author: acmilone@das.inpe.br},
Ronaldo da Silva$^a$,
Anne Sansom$^b$, \\
and Patricia S\'anchez-Bl\'azquez$^c$
}
\smallskip

{\small
$^a$ Instituto Nacional de Pesquisas Espaciais, S. J. dos Campos, Brazil\\
$^b$ University of Central Lancashire, Preston, UK\\
$^c$ Universidad Aut\'onoma de Madrid, Madrid, Spain\\
}

{\footnotesize Received on September 30, 2011 / accepted on December 22, 2012}

\end{center}


\begin{abstract}
We report on two distinct computational approaches
to self-consis\-tently measure photospheric properties of large samples of stars.
Both procedures consist of a set of several semi-integrated tasks based on shell and Python scripts,
which efficiently run either our own codes or open source software commonly adopted by the astronomical community.
One approach aims to derive the main stellar photospheric parameters and abundances of a few elements
by analysing high-resolution spectra from a given public library homogeneously cons\-tructed.
The other one is applied to recover the abundance of a single element in stars with known photospheric parameters
by using mid-resolution spectra from another open homogeneous database
and calibrating derived abundances. 
Both semi-automated computational approaches provide homogeneity and objectivity to every step of the process
and represent a fast way to reach partial and final results
as well as to estimate measurement errors,
making possible to systema\-tically evaluate and improve the distinct steps.

\bigskip

\end{abstract}

\textbf{1. INTRODUCTION}
\bigskip
\bigskip

The physical, chemical and dynamical characterization of a stellar photosphere is made by analysing its spectrum.
To achieve this astronomers need to measure the intensity and/or shape of many atomic and molecular
absorption lines in high-quality optical spectrum at high resolution
(resolving power $R$ $\equiv$ $\lambda$/$\Delta\lambda$ $>$ 10\,000, $\Delta\lambda$ = $FWHM$).
However, adopting mid-resolution spectra (1\,000 $\leq$ $R$ $\leq$ 10\,000)
represents an alternative either to obtain the photospheric temperature, gravity and metallicity on large surveys
when associated with photometric measurements and comparisons against other data sets ($R$ $\approx$ 2\,000)
\cite {Lee2008}
or to recover multi-elemental abundances for previously characterized stars on smaller samples ($R$ $\approx$ 6\,000)
\cite {Kirby2009}.

The first step of any stellar spectroscopic analysis at any resolution is the spectral continuum normalization.
The overall continuum shape primarily depends on the photospheric temperature
but also shows particularities as a function of the wavelength due to the total local opacity
(sum of all processes of light absorption and scattering
integrated along the photosphere at a given frequency).
The second step is to measure the intensity of each atomic absorption feature,
for which the profile can be represented by the sum of one or more individual lines.
Both steps are very time-consuming tasks that also include some level of subjectivity in the selection of continuum windows and absorption features.
It is normally performed using known public programs for analysis of astronomical data
such as the IRAF\footnote{
Image Reduction and Analysis Facility, IRAF, is distributed by the National Optical Astronomy Observatories, USA.}
multi-platform software and others developed by the astronomers themselves that make the features measurements faster (e.g. ARES
\cite {Sousa2007}).
Besides the measurements of a myriad of atomic lines,
the other inputs for the spectroscopic analysis are the physical structural model of the photosphere
and the physical parameters of a representative list of atomic
and molecular lines to be taken into account in the theoretical spectrum computation,
which is the following step and also consumes a lot of time.

In this paper we describe two semi-automated approaches developed to alleviate those issues and to objectively analyse large stellar spectral data sets.
In the first one, the fundamental photospheric parameters and abundances of several elements are self-consistently derived
by analysing high-resolution spectra of stars selected from the ELODIE database
\cite {Moultaka2004}.
The other computational approach addresses the measurement of magnesium abundance in stars of the MILES spectral library
\cite {SanchezBlazquez2006}
by inspecting the own MILES' mid-resolution spectra.
The photospheric parameters of MILES' stars are previously known and
were homogenized onto a uniform system from measurements of several published works
\cite {Cenarro2007}.
The procedure presented here includes also the calibration of resulting mid-resolution Mg abundances
against homogeneously scaled high-resolution measurements.

\bigskip
\bigskip
\textbf{2. DETAILED PHOTOSPHERIC CHARACTERIZATION \linebreak 
AT HIGH SPECTRAL RESOLUTION}
\bigskip
\bigskip

The complete characterization of a stellar photosphere
depends on the computation of a realistic model atmosphere,
which is described by four fundamental parameters:
the effective temperature $T_{\rm eff}$,
the metal content traced by the iron abundance
that is usually represented in a logarithm scale relative to the solar pattern [Fe/H]\footnote{
[Fe/H] = $\log(\epsilon({\rm Fe})/\epsilon({\rm H}))_{\star}$ $-$ $\log(\epsilon({\rm Fe})/\epsilon({\rm H}))_{\odot}$
where $\log{\epsilon({\rm Fe})}$ = $\log(n({\rm Fe})/n({\rm H}))$ + 12, $\log{\epsilon({\rm H})}$ = 12 and
$n$(Fe) and $n$(H) are the number densities (volumetric or columnar) of iron and hydrogen atoms respectively.},
the logarithmic surface gravity $\log{g}$, and the micro-turbulence velocity $\xi$.
High-resolution measurements of many optical lines of neutral and first ionized states of iron atoms (Fe~I and Fe~II respectively)
have been historically adopted for this purpose.

In the work of
Da Silva et al. (2011)
\cite {DaSilva2011}
an automated computational procedure was developed to analyse high-resolution spectra ($R$ = 42\,000)
of 172 solar-like stars collected from the ELODIE database.
The method adopts the Sun's parameters as an input set
($T_{\rm eff}$ = 5777~K, [Fe/H] = 0.0 dex, $\log{g}$ = 4.44, $\xi$ = 1.0~km.s$^{-1}$)
to find the best solution for the model atmosphere through an iterative process described next.
This automated procedure was also applied by Da Silva \& Milone (in prep.)
to a wider sample of ELODIE's solar-type stars.

The abundance yielded by distinct spectral lines of an element E ($\log{\epsilon({\rm E})_{\star}}$ or [E/H])
must not be dependent on the excitation potential ($\chi$) and equi\-valent width ($EW$).
The excitation potential is the lowest energy level of the electronic transition,
and the equivalent width quantifies the electromagnetic flux absorbed by the transition over the emergent photospheric spectrum
($EW$ $\equiv$ $\int (1-F(\lambda)_{l}/F(\lambda)_{c}).d\lambda$,
where $F_{l}$ and $F_{c}$ are the line and continuum flux respectively).
Also, the lines of neutral and ionised species from the same element must provide a unique abundance.
$T_{\rm eff}$ is normally computed through the excitation equilibrium of Fe~I
by removing any dependence in the $\log{\epsilon({\rm Fe})}$ vs. $\chi$ diagram.
Additionally, by removing any spurious abundance dependence on $EW$, $\xi$ is estimated.
Surface gravity is computed through the ionisation equilibrium between the Fe~I and Fe~II species,
and finally [Fe/H] is derived from the $EW$ measurements of Fe~I lines.

\begin{figure}[]
\begin{center}
\includegraphics[angle=90, height=6.6cm]{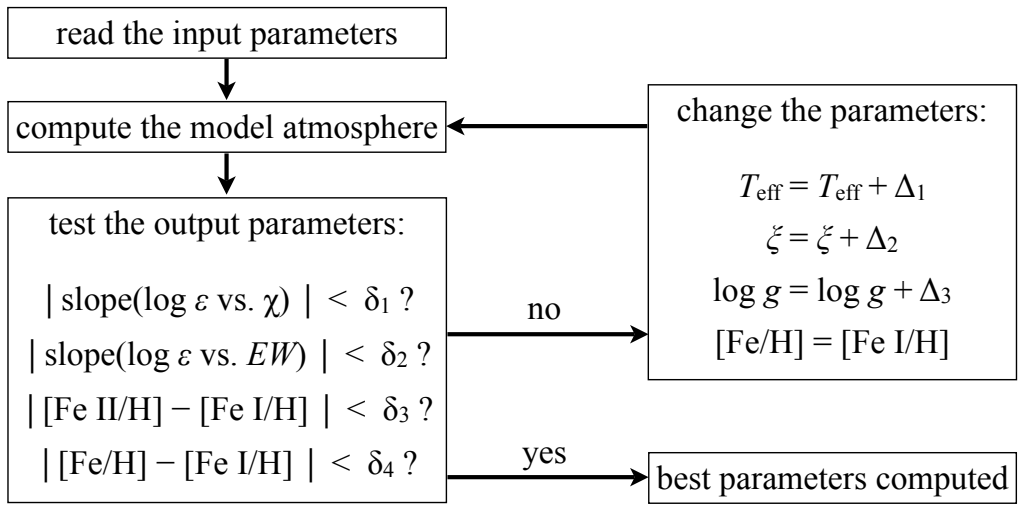}
\caption{Flux diagram for the spectroscopic analysis to determine the stellar photospheric parameters at high resolution.}
\end{center}
\end{figure}

Figure 1 illustrates through a flux diagram how the photospheric para\-meters are determined,
where $\delta_1$, $\delta_2$, $\delta_3$, and $\delta_4$ are arbitrary constants as small as one wishes.
If at least one of the first three conditions is not satisfied, then $T_{\rm eff}$, $\xi$,
and/or $\log{g}$ are changed by a given step (respectively $\Delta_1$, $\Delta_2$ and $\Delta_3$).
In the fourth condition, the value of [Fe/H] used as input is compared to the one provided by Fe~I lines
and, if they do not agree within $\delta_4$, the code defines [Fe/H] = [Fe~I/H].
Therefore, the code iteratively executes several cycles until these four conditions are satisfied together.
We tested the method using different sets of input parameters for distinct stars
and the same solution is always reached for each star considering their uncertainties.

The computational method for obtaining the photospheric parameters is based on a Python script,
which incorporates:
($i$) a Fortran code for interpolation in a grid of model atmospheres
\cite {Kurucz1993};
($ii$) the execution of a spectral synthesis code (MOOG 
\cite {Sneden2002})
applied to the interpolated model by computing the theoretical predictions of $EW$ for a set of Fe~I and Fe~II lines;
and $(iii)$ the simultaneous convergence conditions through an iterative analysis of the input/output parameters.

\begin{figure}[]
\begin{center}
\includegraphics[angle=0, height=6.2cm]{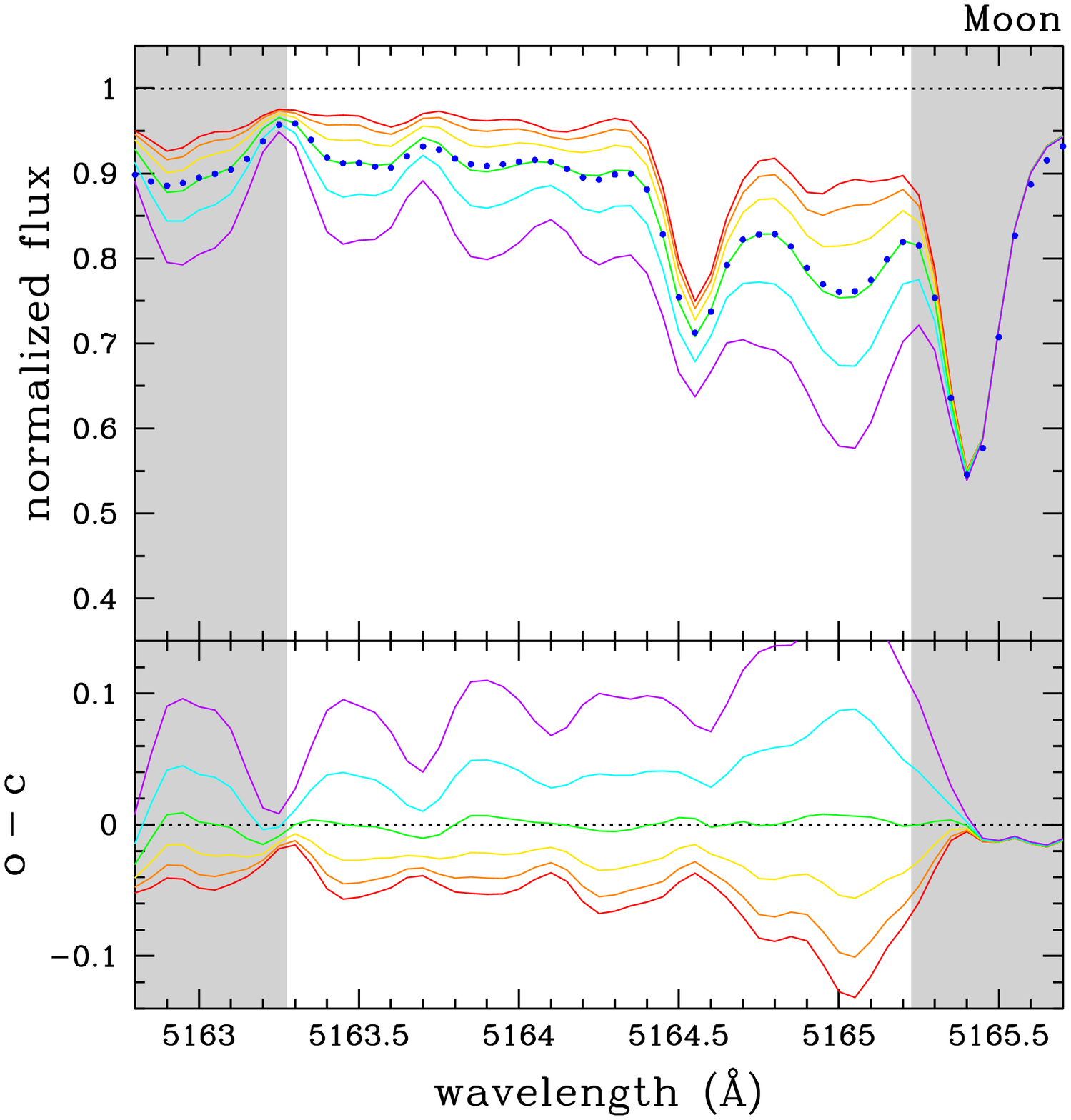}
\includegraphics[angle=0, height=6.2cm]{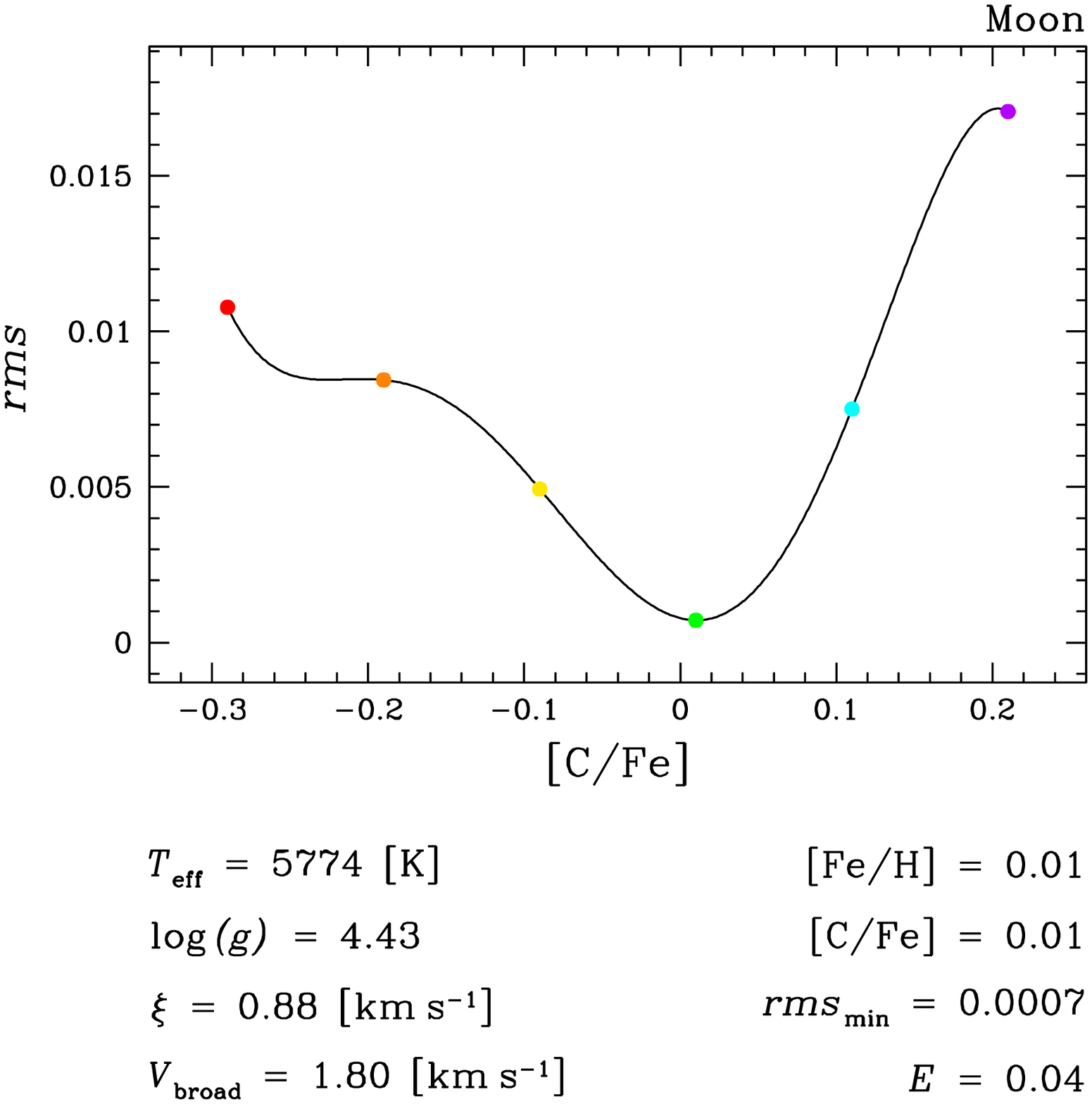}
\caption{Example of a spectral synthesis at high resolution of the C$_{2}$ band feature at $\lambda$5165{\AA}
(left panel, from Fig. 3 of Da Silva et al. (2011))
and the diagnostic plot $rms$ of theoretical and observed spectral comparisons versus [C/Fe] (right panel).
The observed spectrum (blue points in the left panel) is the sunlight reflected by the Moon.
The flux continuum normalization is checked in other plots.}
\end{center}
\end{figure}

The semi-automated element abundance determinations performed in Da Silva et al. (2011)
and in Da Silva \& Milone (in prep.) are based on $EWs$ of selected atomic lines (Na, Mg, Si, Ca, Ti, V, Mn, Fe, Ni, Cu, and Ba)
as well as on spectral synthesis of C$_{2}$, CN, C I, O I, and Na I features,
where both techniques are applied with the MOOG code.
The procedures were incorporated into a Python script, which includes:
($i$) the computation of the element abundances for each star
based on their photospheric parameters and $EWs$,
fixing the solar $gf$ values
(these were revised to fit the solar observed $EWs$ because the Sun is taken as a reference);
($ii$) the computation of synthetic spectra for six different values of [E/Fe]
conside\-ring the photospheric parameters, the continuum opacity contribution, and the line-broadening corrections;
($iii$) a vertical shift of the synthetic spectra based on continuum windows;
($iv$) the computation of the $rms$ deviation between synthetic and observed spectra
within selected regions of atomic and molecular features; and
($v$) an iterative search for the smallest $rms$ value,
which is done by changing the six abundance values until finding a minimum in the $rms$ vs. [E/Fe] diagram.

Figure 2 presents an example of the best spectral synthesis of a C$_{2}$ feature
that was reached through this computational procedure.
The resulting carbon abundance is derived using a polynomial fit of the $rms$ deviation
between the synthetic and observed spectra as a function of the [C/Fe] abundance ratio
as illustrated in a diagnostic plot employed in the process (shown also in Fig. 2).

An error analysis for the derived photospheric parameters and element abundances was also performed in an automated way.
Basically the stellar parameter uncertainties are iteratively computed
through a Python script following an analogue computational procedure
as that applied to determine the stellar parameters themselves (see Fig. 1).
The abundance errors are estimated considering the propagation of atmospheric parameter errors.
They are automatically calculated adopting another Python script.

\bigskip
\bigskip
\textbf{3. PHOTOSPHERIC Mg ABUNDANCE MEASURED \linebreak 
AT MEDIUM SPECTRAL RESOLUTION}
\bigskip
\bigskip

In the computational approach that semi-automatically runs the mid-resolution spectroscopic analysis ($R$ $\approx$ 2\,200)
carried out by Milone et al. (2011)
\cite {Milone2011},
the photospheric parameters were not determined
because they were already known.
The objective in that work was to measure the magnesium abundance in the stars of the MILES optical spectral library.
For this purpose we first computed a large number of model atmospheres
to represent the physical-chemical conditions of each photosphere.
Linear interpolations of the MARCS model atmospheres
\cite {Gustafsson2008}
were applied using a public code
\cite {Masseron2006},
which is automatically manipulated through a C shell script.
This was improved by Milone et al. (2011)
to incorporate some specific procedures such as
($i$) reading the input list of MILES stars together with their parameters,
($ii$) definition of eight input models to be read from the MARCS grid,
($iii$) inclusion of a table of element abundances and molecular species into each model file, and
($iv$) transformation of the output model files into a format readable by the MOOG code.
A global picture of the whole semi-automated computational approach described in this section is
illustrated in Figure 3.

\begin{figure}[]
\begin{center}
\includegraphics[angle=0, width=10.2cm]{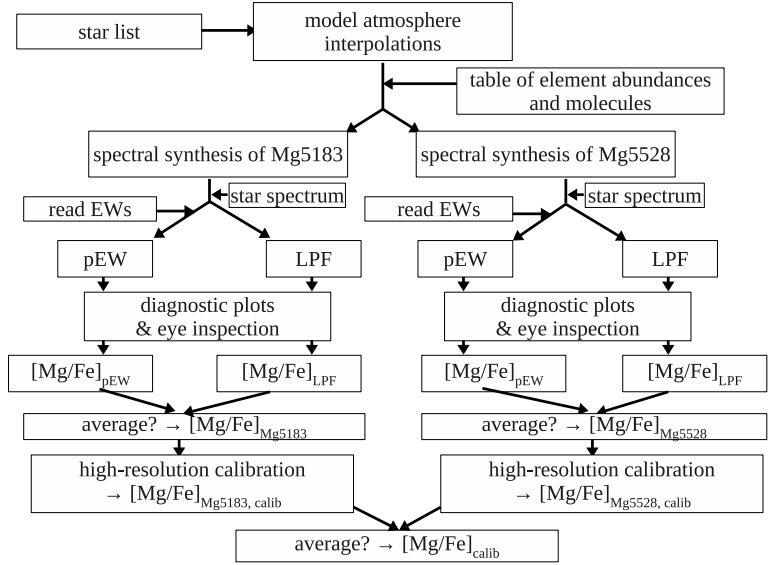}
\caption{Flux diagram for the spectroscopic analysis to derive the [Mg/Fe] abundance ratio at medium resolution.}
\end{center}
\end{figure}

After the model atmospheres have been built up through interpolations,
five synthetic spectra had to be computed at medium resolution in two different regions for each MILES' star inside the MARCS grid.
One Mg~I atomic feature was analysed per spectrum region:
the strongest line of Mg b triplet at $\lambda$5183.6{\AA}, and the line at $\lambda$5528.4{\AA}.
Two element abundance determination methods were employed:
the pseudo-equivalent width (pEW) and the line profile fit (LPF),
both applied to each feature passband adopting two pseudo-continuum windows.
Figure 4 shows an example of spectral synthesis for the Mg5183 feature
in association with both methods applied to measure the Mg abundance.
843 MILES stars were analysed and 8,430 theoretical spectra had to be computed through an semi-automated process
based on a very simple Expect shell script that successively runs the MOOG software thousands of times. 
Every synthetic spectrum incorporates a Gaussian-like instrumental broadening to reach the MILES' resolution.
Their wavelength scales were re-binned to exactly agree with the sampling of the MILES spectra at each observed wavelength bin.
Additionally, the wavelength scales of MILES spectra were carefully shifted to the rest frame to match the model scale.
To guarantee a reliable matching between the flux scales of model and empirical spectra,
the latter ones had their pseudo-continuum flux locally normalized in both Mg~I feature regions.
All these steps were executed by C shell scripts associated to IRAF tasks
(e.g. FXCOR of the NOAO Optical Astronomy Package
for spectral cross-correlations and to obtain pixel fractional corrections for reaching the rest wavelength scale).

\begin{figure}[]
\begin{center}
\includegraphics[angle=-90, width=11.4cm]{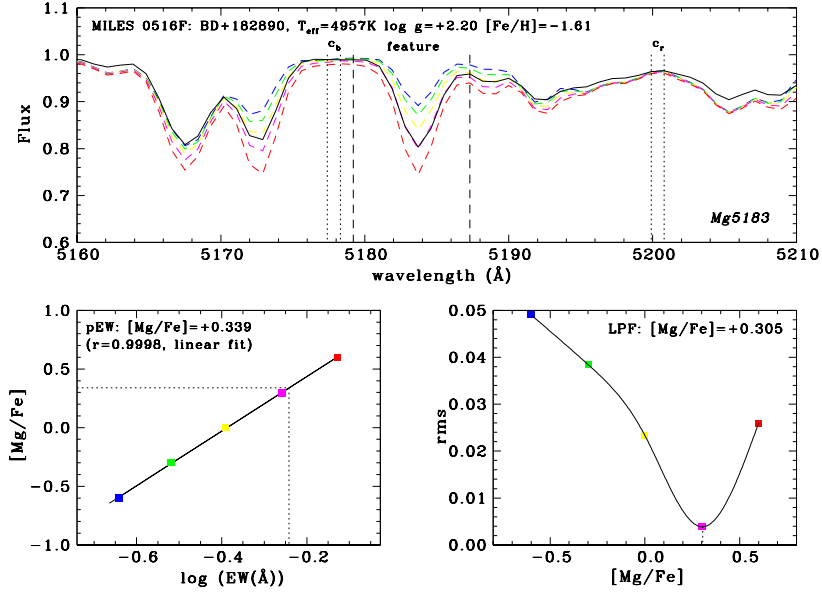}
\caption{Example of diagnostic plots, adapted from Milone et al. (2011),
for the spectral synthesis of Mg5183 feature (top panel)
and the correspondent methods employed to determine [Mg/Fe]
(pseudo-equivalent width in the left bottom panel and line profile fit in the right one).}
\end{center}
\end{figure}

The other steps of the computational approach for recovering [Mg/Fe] abundance ratios
at medium resolution applied by Milone et al. (2011)
can be summarize as follows (see also Fig. 3).
\begin{enumerate}
\item Automated measuring of $EW$ of two Mg features over all empirical and theoretical spectra
(by running the LECTOR public Fortran code, www.iac.es/galeria/vazdekis/SOFTWARE/, through C shell scripts).
\item Application of the pEW and LPF methods based on several concatenated C shell scripts,
together with a macro developed by us using an interactive plotting code (SM
\cite {Lupton2000}),
which employs a flexible command language.
This step designs diagnostics plots of the spectral synthesis of both Mg features for each star (example in Fig. 4).
\item For each Mg feature, computation of the mean [Mg/Fe] value of those obtained with the two methods
after inspection by eye in the spectral synthesis plots.
The average is only computed when both methods show reliable results
\footnote{
in some cases, only one method gives a reliable value while in others both methods fail.
This gives [Mg/Fe]$_{\rm feature}$}.
\item Linear calibration of [Mg/Fe]$_{\rm feature}$ to a uniform scale based on our own Fortran routines,
separately for each feature, via comparison with a control sample of MILES stars
(for which the abundance ratios were obtained from published high-resolution analyses
and previously calibrated onto a homogeneous scale),
in order to provide [Mg/Fe]$_{\rm feature}^{\rm calib}$.
\item Simple averaging of [Mg/Fe]$_{\rm feature}^{\rm calib}$ combined when possible, 
in order to derive the final calibrated [Mg/Fe]$_{\rm calib}$ for each star (Fortran routines from ourselves).
\end{enumerate}

An error analysis was also performed along the whole process (step by step based on own codes and graphics)
in order to estimate the intrinsic uncertainties from each abundance determination method,
the uncertainties due to the error propagation from photospheric parameters,
the effective internal errors of both [Mg/Fe]$_{\rm feature}$, 
and the systematic or external errors of final calibrated abundance ratios
[Mg/Fe]$_{\rm feature}^{\rm calib}$ and [Mg/Fe]$_{\rm calib}$.

\bigskip
\bigskip
\textbf{4. CONCLUSIONS}
\bigskip
\bigskip

On the one hand, a stellar spectroscopic analysis at high resolution commonly offers better results
than an analysis at medium resolution for determining photospheric parameters and element abundances.
For instance, the typical internal uncertainties of $T_{\rm eff}$, [Fe/H], $\log{g}$, $\xi$ and [C/Fe] in Da Silva et al. (2011)
are, respectively, 70~K, 0.06 dex, 0.2, 0.07~km.s$^{-1}$ and 0.04 dex,
while the external error of [Mg/Fe]$_{\rm calib}$ in Milone et al. (2011)
ranges from 0.10 to 0.15 dex according to the Mg feature adopted (average value of 0.12 dex).
On the other hand, a medium-resolution analysis that is partially or fully automated
can provide astronomers nowadays with an efficient tool to analyse large samples
as done by Milone et al. (2011)
to recover magnesium abundances in the stars of MILES library
as well as those performed by
\cite {Lee2008} and
\cite {Lee2011}
to respectively measure the stellar parameters and abundances over the SDSS/SEGUE wide survey.
In
\cite {Lee2008} and
\cite {Lee2011},
other techniques were also employed in automatic way
such as the SDSS photometry, empirical-theoretical fitting of full spectrum, and line strength indices
to help recovering the photospheric properties with more precision.
The reached internal errors of $T_{\rm eff}$, [Fe/H], $\log{g}$ and [E/Fe] are typically 70~K, 0.07 dex, 0.18 and smaller than 0.1 dex,
while the external errors of them are 130~K, 0.11 dex, 0.21 and 0.1 dex.
It is out of the scope of the current work
to compare other methods found in the literature and their computational procedures
against all those applied here.
The techniques applied by us at high and medium resolution are fully self-consistent
and the computational procedures developed here have represented an objective solution to manipulate them with reliability and efficiency.

The classical human-interactive approach performs individual analysis of each star
carrying out several steps one by one in a tedious iterative and frequently subjective way
(e.g. spectral synthesis of a given feature per time).
The major distinctions of our semi-automated spectroscopic analyses in comparison with the classical approach are:
($a$) manipulation of several scripts and software written in different languages, and
($b$) automated execution of concatenated steps of the whole process.
Their main goals are:
($i$) to guarantee homogeneity and a high level of objectivity in the application of each step of the process,
($ii$) to reach partial and final results in a fast way making possible to evaluate the separated steps,
($iii$) to rapidly estimate measurement errors step by step, and
($iv$) to easily repeat the whole process or its distinct steps as many times as necessary
to correct minor mistakes and improve parts of the process.
Although the gain in CPU time is actually not improved,
we have estimated that the gain in man-power time compa\-ratively with the classical approach
can reach to around a factor of 100 in the determination of the photospheric parameters,
and about a factor of 10 in the application of spectral synthesis to reproduce line profiles and obtain element abundances.

We intend to make both semi-automated computational approaches pu\-blicly available
when their tools are consistently integrated and documented in order to be easily installed and manipulated in any plataform.

\bigskip

{\bf ACKNOWLEDGMENTS: AM thanks the Brazilian foundation CAPES (grant BEX 2895/07-2). RDS thanks the institutional programme PCI/MCTI/INPE (grant PCI-DA 300422/2011-3).

\end{document}